\documentclass[amsmath, amssymb, aps, twocolumn]{revtex4-1}
\usepackage{graphicx}
\usepackage{amsmath}
\usepackage{bm}
\usepackage{mathrsfs}
\usepackage{booktabs}
\usepackage{multirow}
\usepackage{tabularx}

\begin{document}

\title{Robust scheme for magnetotransport analysis in topological insulators}

\author{G. Eguchi}
\email{eguchi@ifp.tuwien.ac.at}
\author{S. Paschen}

\affiliation{Institute of Solid State Physics, Vienna University of Technology, Wiedner Hauptstrasse 8-10, 1040 Vienna, Austria}

\date{\today}

\begin{abstract}
The recent excitement about Dirac and Weyl fermion systems has renewed interest in magnetotransport properties of multi-carrier systems. However, the complexity of their analysis, even in the simplest two-carrier case, has hampered a good understanding of the underlying phenomena. Here we propose a new analysis scheme for two independent conduction channels, that strongly reduces previous ambiguities and allows to draw robust conclusions. This is demonstrated explicitly for the example of three-dimensional topological insulators. Their temperature and gate voltage-dependent Hall coefficient and transverse magnetoresistance behavior, including the phenomenon of huge linear transverse magnetoresistance, can be traced back to two conduction channels, with fully determined carrier concentrations and mobilities. We further derive an upper limit for the transverse magnetoresistance. Its violation implies field dependences in the electronic band structure or scattering processes, or the presence of more than two effective carrier types. Remarkably, none of the three-dimensional topological insulators or semimetals with particularly large transverse magnetoresistance violates this limit.

\end{abstract}

\maketitle
\section{Introduction}
Topologically non-trivial insulators and semimetals continue to be of great interest~\cite{RevModPhys.90.015001,Liu2018,PhysRevB.97.245101}, both to advance the fundamental understanding of topological matter and to pave the way for new applications. 
Indications for their existence come mostly from surface-sensitive probes such as angle-resolved photoemission spectroscopy and scanning probe microscopy~\cite{nphys1270,PhysRevLett.105.146801,PhysRevLett.105.136802,nmat3990,Xue1501092,PhysRevB.95.125126}, whereas evidence from bulk probes such as (magneto)transport is more circumstantial~\cite{PhysRevB.81.241301,PhysRevB.82.241306,PhysRevB.91.041203,srep04859,nphys3372}. This is due to the fact that topologically trivial electronic states typically coexist with the Dirac or Weyl states. To disentangle these two components, the two-band Drude model~\cite{PhysRevB.82.241306,science.1189792,acsnano.5b00102,nmat4143,nphys3372,PhysRevB.95.045123} has been used. However, the large number of open parameters makes these analyses extremely unreliable~\cite{Colin}, which hampers progress.

Here we propose a new scheme for such analyses, that does not require any {\em ad hoc} assumption on the charge carrier types, concentrations or mobilities, and instead determines all these quantities explicitly. It thus allows to reveal the physical origin of characteristic phenomena such as the sign inversion of the Hall coefficient and the huge linear transverse magnetoresistance (TrMR)~\cite{srep04859,nphys3372}. 
In addition the Fermi level as well as the upper limit for TrMR can be determined as functions of the Hall mobility and field.

\section{Formulation}
We start by describing the differences between the common two-carrier analysis and our new scheme. In the former, the resistance $R_{xx}(B)$ and the Hall resistance $R_{xy}(B)$, where $B$ is the magnetic field, are characterized by four free parameters (the charge carrier concentrations $n_1$ and $n_2$, and the mobilities $\mu_1$ and $\mu_2$) and two constant parameters (the charge carrier types $q_1$, $q_2 = \pm e$, where $e$ is the elementary charge) that need to be anticipated. 
Following the usual notation, $n_i$ is positive for both electrons and holes, whereas $\mu_i$ is negative for electrons and positive for holes. 
In the new scheme, no {\em ad hoc} assumption has to be made on the charge carrier type. In addition, there are only two free parameters, namely the relative charge carrier concentration 
\begin{equation}
N \equiv \frac{n_1-n_2}{n_1+n_2} = \frac{n_1-n_2}{n_+} \quad ,
\label{eq0a}
\end{equation}
where $n_+ = n_1+n_2$ is the total charge carrier concentration,  and the mobility difference 
\begin{equation}
M \equiv \frac{\mu_1-\mu_2}{\mu_1+\mu_2} \quad ,
\label{eq1a}
\end{equation}
respectively. Two further parameters, the Hall coefficient 
\begin{equation}
R_{\rm{H}} \equiv \lim_{B \to 0} \frac{R_{xy}}{B} \quad ,
\label{eq0c}
\end{equation}
and the Hall mobility 
\begin{equation}
\mu_{\rm{H}} \equiv  \lim_{B \to 0}\frac{R_{xy}}{R_{xx}B} \quad 
\label{eq0d}
\end{equation}
can be directly read off the data.
$N$ and $M$ can thus be determined with minimal ambiguity. 
$q_1$ and $q_2$ are determined as functions of $N$, $M$, and $\mu_{\rm{H}}$ (see Sect.\,A, B of the Appendix for a complete description). 
Several examples where imprecise or even totally erroneous results were obtained with the standard scheme and where our new scheme results in new insights are presented in Sect.\,D of the Appendix. 
The merit of the new scheme is that it reduces the ambiguity in the derived parameter values drastically and that the risk of overlooking unexpected carrier types~\cite{PhysRevB.95.045123,nmat4143}. It thus allows to draw robust solutions.

\section{Application I: Topological insulators}

\begin{figure}[ht]
\begin{center}
\includegraphics[width=\columnwidth,clip]{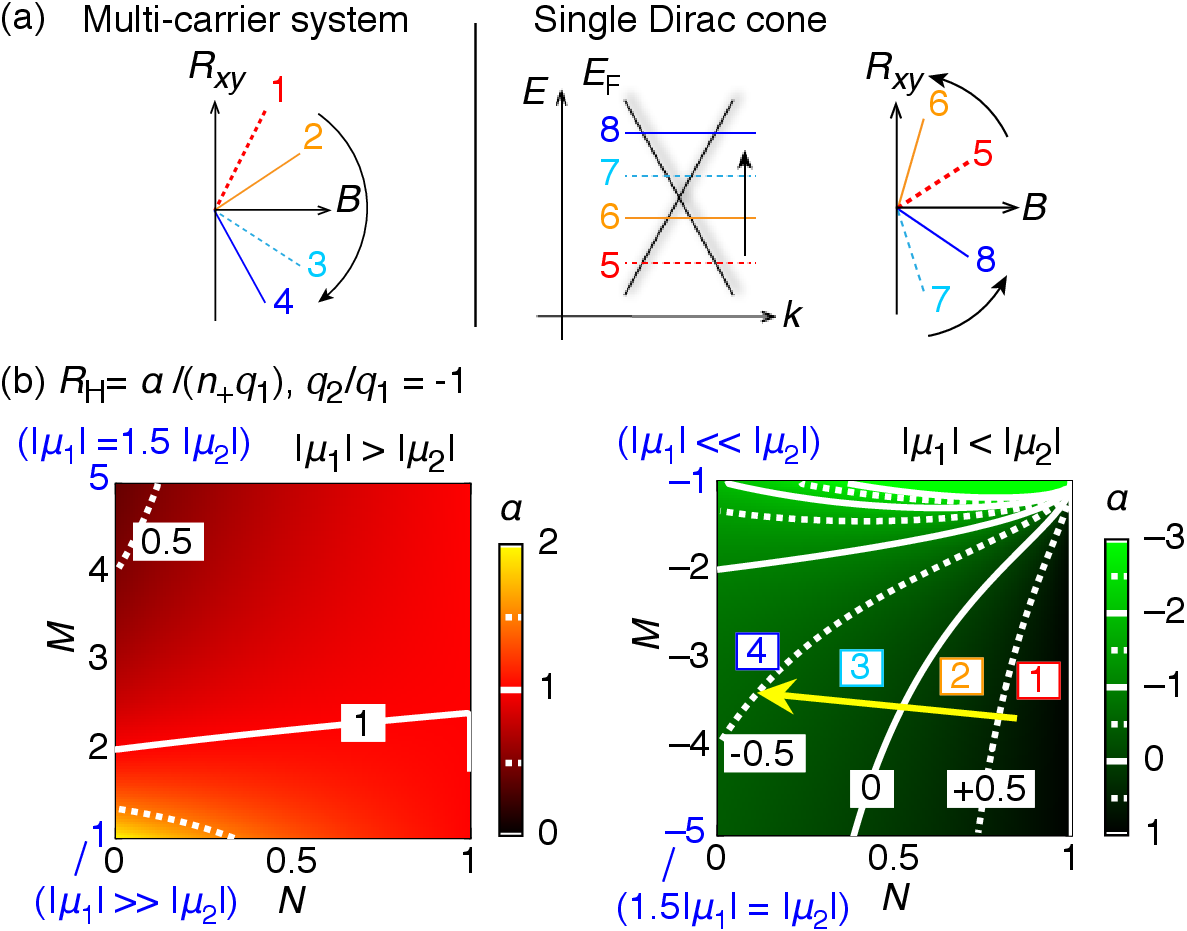}
\caption{(Color online) (a,\,left) Sign inversion of $R_{xy}(B)$ traces (1-4) typically observed in a multi-carrier system as a function of temperature $T$ or gate voltage $V_{\rm{G}}$. 
The slope of $R_{xy}(B)$ varies continuously from 1 to 4. 
(a, right) Electronic band dispersion of a single Dirac cone, where $E$ is the energy and $k$ is the wave vector (left), and corresponding $R_{xy}(B)$ traces at $T=0$ (right) as $E_{\rm{F}}$ is varied across the Dirac point (5-8). 
The sign change of $R_{xy}$ is accompanied by a discontinuity in the slope of $R_{xy}(B)$. 
(b) Contour plots of the Hall factor $\alpha$ for $q_2/q_1=-1$ as functions of $N$ and $M$, for $|\mu_1|>1.5|\mu_2|$ (left) and $1.5|\mu_1|<|\mu_2|$ (right). 
The four situations (1-4) depicted in (a) could arise, for instance, along the arrow, with sign inversion between 2 and 3. 
The main text provides a complete description.
}
\label{fig1}
\end{center}
\end{figure}

Next we show how to understand the sign inversion in $R_{\rm{H}}$, frequently observed in topologically non-trivial materials as a function of temperature $T$ or gate voltage $V_{\rm{G}}$. 
Figure\,\ref{fig1}\,(a, left) depicts the typically observed signature in consecutive (1-4) $R_{xy}(B)$ traces: the slope varies continuously from 1 to 4. By contrast, if the Fermi level $E_{\rm{F}}$ is varied across the Dirac point of a single Dirac cone (5-8 in Fig.\,\ref{fig1}\, (a, right)) at zero temperature ($T=0$) \cite{nature04235}, the sign change of $R_{xy}$ is accompanied by a discontinuity in the slope of $R_{xy}(B)$.

If $n_1$, $\mu_1$, and $q_1$ represent the majority carriers ($n_1 > n_2$) and if charge carriers of opposite sign, i.e., both electrons and holes, are present ($q_2/q_1=-1$), $R_{\rm{H}}$ can be transformed from its original form (Eqn.\,\ref{eq2s1} of the Appendix) to
\begin{equation}
R_{\rm{H}}= \frac{1}{n_+q_1} \alpha \quad , \quad \quad \alpha= \frac{N+2M+NM^2}{(N+M)^2}
\label{eq0a}
\end{equation}
where $\alpha$ is the Hall factor.
The values of the parameters $N$ and $M$, and thus the one of $\alpha$, follow from these definitions as $0<N<1$ and $1<|M|<\infty$. 
Figure\,\ref{fig1}\,(b) shows contour plots of $\alpha$ in the full $N$ range and for $1<M<5$ (left) and $-5<M<-1$ (right). 
Smaller mobility differences would be captured by plots to larger values of $|M|$. 
These are, however, less relevant here because we aim at separating contributions of highly mobile Dirac fermions from those of topologically trivial fermions with much lower mobility.

Our first key result follows directly from these contour plots. A sign change of $\alpha$ and thus of $R_{\rm{H}}$ occurs only in the right panel of Fig.\,\ref{fig1}\,(b) i.e., if the majority carriers have lower mobility than the minority carriers. 
In experiments on putative three-dimensional topological insulators (3D-TIs), where the observed sign inversion in $R_{\rm{H}}$ (Sect.\,D of the Appendix) was taken as evidence for the presence of Dirac surface states~\cite{science.1189792,acsnano.5b00102,srep04859}, the more mobile Dirac fermions must thus have been the minority carriers and topologically trivial charge carriers of lower mobility, most likely associated with residual bulk states the majority carriers. 
To illustrate this further, temperature and gate voltage tuning are given as two concrete examples in what follows.

\begin{figure}[t]
\begin{center}
\includegraphics[width=\columnwidth,clip]{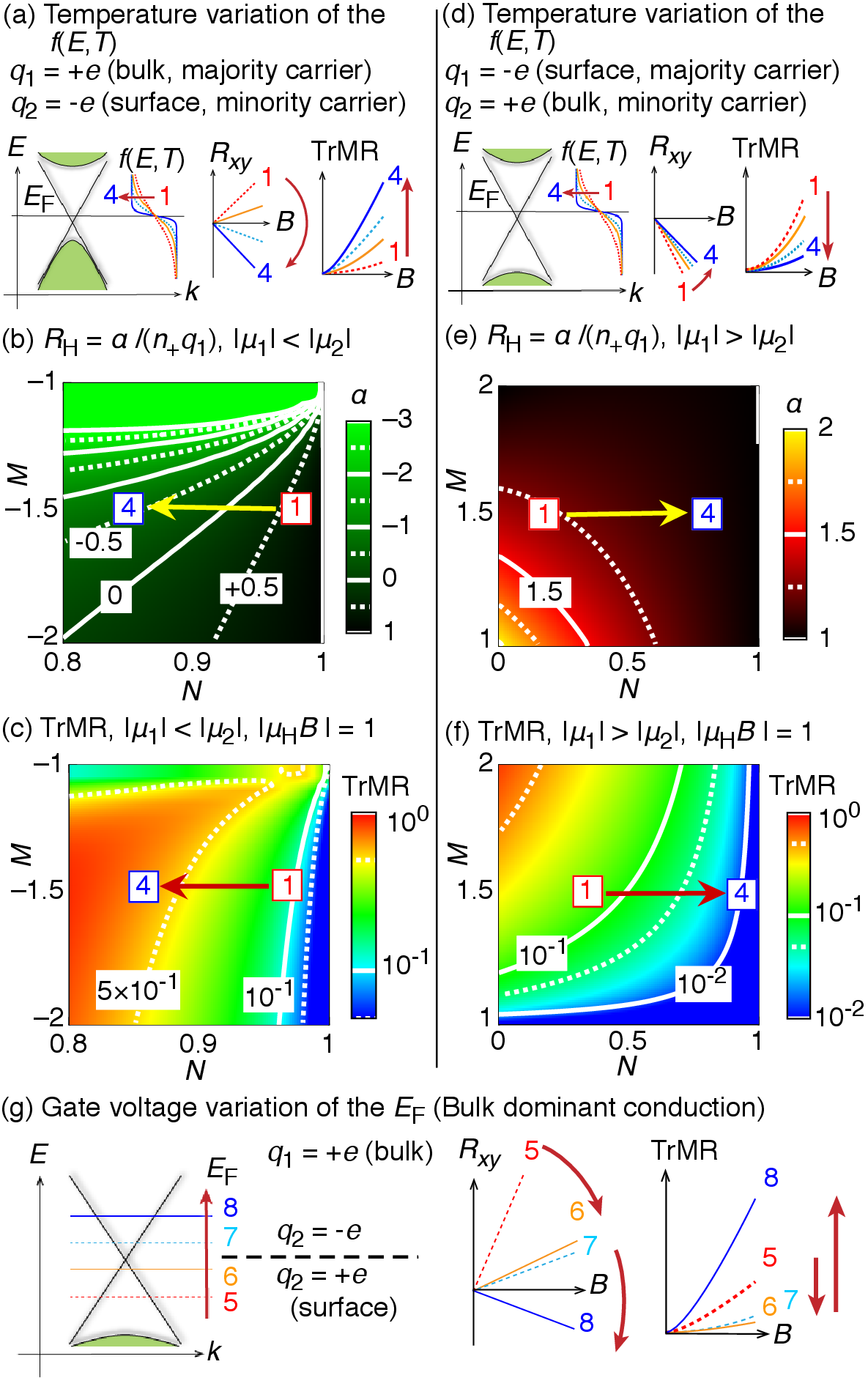}
\caption{
(Color online) (a,\,d,\,left) Sketch of the electronic band dispersion of a 3D-TI and the $f(E,T)$ for four different temperatures (1: highest $T$, 4: lowest $T$). (a,\,d,\,center,\,right) Corresponding $R_{xy}$ and TrMR traces at each temperature. (b,\,e) Contour plots of the $\alpha(N, M)$. (c,\,f) Contour plots of the ${\rm{TrMR}}(N,\,M, |\mu_{\rm{H}}B|=1$). (g,\,left) Sketch of the electronic band dispersion near the Dirac point and  a gate voltage tuning of the $E_{\rm{F}}$ (5: lowest  $E_{\rm{F}}$, 8: highest $E_{\rm{F}}$). (g,\,center,\,right) Corresponding $R_{xy}$ and TrMR traces. See Sect.\,C of the Appendix for the contour plots. The main text provides a complete description.
}
\label{fig2}
\end{center}
\end{figure}

Firstly, the experimentally observed $R_{\rm{H}}$ sign inversion as a function of temperature can be understood by taking the temperature dependence of the Fermi distribution function $f(E,T)$ into account. 
Figure~\ref{fig2}\,(a,\,left) shows a sketch of the electronic band structure of a 3D-TI, for the situation where $E_{\rm{F}}$ lies slightly above the Dirac point. The Dirac fermions are thus electron like. 
The temperature dependence of $f(E,T)$ is also sketched, for different temperatures decreasing from 1 to 4. 
At high temperatures (1), the majority carriers ($n_1$) are thermally excited bulk holes. 
With decreasing temperature ($1\rightarrow 4$), $n_1$ decreases exponentially, resulting in a decrease of $n_+$ and $N$. 
Minor variations are also expected for $n_2$, $\mu_1$, and $\mu_2$, but they are neglected here for simplicity. 
To extract information on the system at the sign inversion of $R_{\rm{H}}$, we replot a section of the $\alpha(N,\,M)$ contour plot of Fig.\,\ref{fig1}\,(b,\,right) around $M=-1.5$ ($|\mu_2/\mu_1|=5$), a situation considered realistic for experimentally studied 3D-TIs, in Fig.\,\ref{fig2}\,(b). 
Upon lowering the temperature ($1\rightarrow 4$), sign inversion ($\alpha = 0$) occurs at $N=0.923$, corresponding to only 4\% of surface carriers ($n_2=0.04 n_1$). 
Larger mobility differences (smaller negative $M$ values, towards the top of Fig.\,\ref{fig2}\,(b)) correspond to even smaller fractions of surface carriers.

Secondly, gate voltage tuning can be mimicked by a variation of $E_{\rm{F}}$ around the Dirac point (Fig.\,\ref{fig2}\,(g,\,left), $E_{\rm{F}}$ increases from 5 to 8).  
If temperature is not too low, the above situation with minority surface carriers ($n_2$) is still relevant here because of the very small density of states of Dirac particles near the Dirac point. The $R_{xy}(B)$ traces near sign inversion are shown in Fig.\,\ref{fig2}\,(g,\,center) ($\alpha(N,\,M)$ contour plots are given in Sect.\,C of the Appendix). 
For the typical mobility ratio $|\mu_2/\mu_1|=5$ ($M=-0.667$ for $q_2/q_1=1$ and $M=-1.5$ for $q_2/q_1=-1$) also considered above, sign inversion occurs at $N=0.923$, corresponding to 4\% of surface carriers, similar to the case of temperature tuning.

Robust information can also be extracted if $R_{\rm{H}}(T)$ reveals no sign inversion. 
Such a situation may arise in a bulk-insulating 3D-TI where surface carriers are the majority carriers (Fig.\,\ref{fig2}\,(d)). Here, the exponential decrease of $n_2$ with decreasing temperature ($1\rightarrow 4$) results in only a small decrease of $n_+$ and a small increase of the $N$ (Fig.\,\ref{fig2}\,(e), arrow assumes again a mobility ratio $|\mu_1/\mu_2|=5$ ($M=1.5$)). 
Such a minor effect on the $R_{xy}(B)$ isotherms can, on its own, hardly be taken as strong evidence for the detection of Dirac fermions. 
However, in conjecture with transverse magnetoresistance measurements, strong conclusions can be drawn, as detailed in what follows.

The transverse magnetoresistance ${\rm{TrMR}}\equiv [R_{xx}(B)-R_{xx}(0)]/R_{xx}(0)$ of a two-carrier system with $q_2/q_1=-1$ can be transformed from its original form (Eqn.\,\ref{eq7s0} of the Appendix) into
\begin{equation}
{\rm{TrMR}} = \frac{(N^2-1)M^2(1-M^2)(\mu_{\rm{H}}B)^2}{(2M+N+NM^2)^2+N^2(1-M^2)^2(\mu_{\rm{H}}B)^2} \quad.
\label{eq2b}
\end{equation}
As in the case of $\alpha$, the new analysis scheme reveals traces of TrMR as a function of $N$ and $M$, for a fixed $\mu_{\rm{H}}B$ (Fig.\,\ref{fig2}\,(c,\,f), and Fig.\,\ref{sfig1s}\,(c) of the Appendix).
$R_{\rm{H}}(T)$ in a system with surface majority carriers (Fig.\,\ref{fig2}\,(d-f)) is, for the exemplary case of $|\mu_{\rm{H}}B|=1$, accompanied by a monotonic decrease of TrMR$(T)$ with decreasing temperature. By contrast, if surface carriers are the minority carriers (Fig.\,\ref{fig2}\,(a-c)), TrMR$(T)$ should increase with decreasing temperature~\cite{PhysRevB.90.201307}.

The sign inversion in $R_{\rm{H}}(V_{\rm{G}})$ is accompanied by a distinct feature in TrMR: a non-monotonic variation of TrMR$(V_{\rm{G}})$ with a minimum of TrMR at the Dirac point (Fig.\,\ref{fig2}\,(g,\,right); for contour plots see Sect.\,C of the Appendix). 
This insight establishes a new technique to determine the position of the Fermi level~\cite{PhysRevB.91.235117,PhysRevB.93.174428} with respect to the Dirac point in the low magnetic field limit. It is thus complementary to Shubnikov-de Haas experiments~\cite{PhysRevB.82.241306,science.1189792}, which typically require very high sample quality, low temperatures, and large Fermi cross-sections to observe the oscillations. Thus, our new scheme is particularly rewarding for ambient temperature measurements, low-mobility systems, and cases with extremely low carrier concentrations. 

\section{Application II: Transverse magnetoresistance}

\begin{figure*}[t]
\begin{center}
\includegraphics[width=\textwidth,clip]{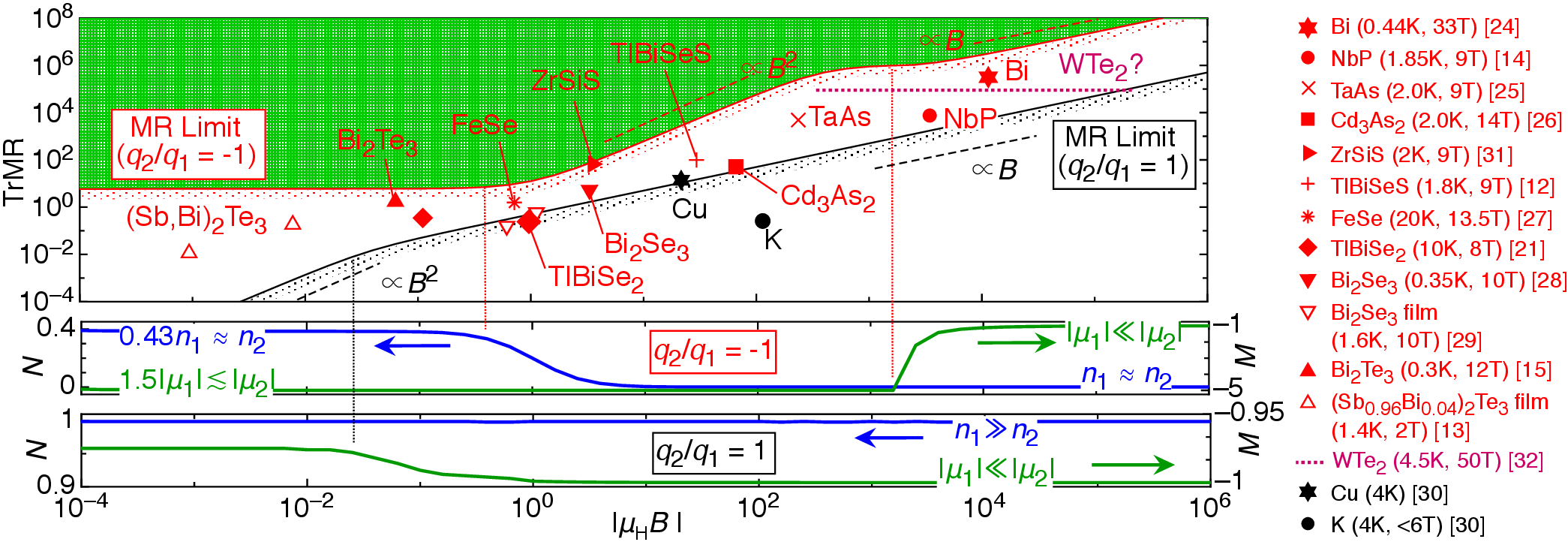}
\caption{(Color online) Upper limit of the TrMR for $q_2/q_1= -1$ (in red) and for $q_2/q_1= 1$ (in black) as a function of $|\mu_{\rm{H}}B|$ (top), and corresponding $N$ and $M$ values (middle and bottom). 
A linear increase of the TrMR limit with $|\mu_{\rm{H}}B|$ occurs if $|\mu_1| \ll |\mu_2|$. 
The largest published TrMR values for Dirac and Weyl fermion systems ($q_2/q_1= -1$, in red) and several pure elements (in black) are also plotted~\cite{Science3171729,nphys3372,PhysRevB.95.085202,PhysRevB.92.081306,PhysRevB.91.041203,PhysRevLett.115.027006,PhysRevB.90.201307,PhysRevB.87.035133,PhysRevLett.109.066803,science.1189792,srep04859,Pippard,PhysRevB.96.045127,nature13763}, with the corresponding $|\mu_{\rm{H}}B|$ values for the same sample batch given in the same reference. 
Remarkably, none of them overshoots the limit.}
\label{fig4}
\end{center}
\end{figure*}

The phenomenon of TrMR has puzzled researchers for a long time. A single-carrier Drude model predicts zero TrMR. 
However, large TrMR values are reported even for the simplest metals such as potassium or copper~\cite{Pippard}. 
This discrepancy was recognized early on~\cite{Ashcroft_Mermin,Colin,Pippard}, and was highlighted again more recently in conjecture with the huge linear TrMR observed in Dirac fermion systems~\cite{PhysRevB.81.241301,PhysRevB.91.041203,srep04859,nphys3372}.

Our two-carrier analysis scheme advances the understanding of this phenomenon by revealing that, for any given $|\mu_{\rm{H}}B|$ value, there is an upper limit to the TrMR (Fig.\,\ref{fig4}). 
Interestingly, for more mobile minority carriers ($|\mu_1| < |\mu_2|$), the TrMR limit increases linearly with $|\mu_{\rm{H}}B|$ for $|\mu_{\rm{H}}B| > 10^3$ if $q_2/q_1=-1$ and for $|\mu_{\rm{H}}B| > 10^{-1}$ if $q_2/q_1=1$. 
The corresponding $N$ and $M$ values are also plotted (middle and lower panel of Fig.\,\ref{fig4}). 
They suggest that a linear increase of the TrMR limit with $|\mu_{\rm{H}}B|$ occurs if $|\mu_1| \ll |\mu_2|$. 
The largest TrMR values for various Dirac or Weyl fermion materials reported in the literature, at the same field together with the corresponding $\mu_{\rm{H}}$ values, are also summarized in Fig.~\ref{fig4}. 
Even bismuth, which exhibits the largest TrMR observed to date~\cite{Science3171729}, does not overshoot the limit. 
For WTe$_2$, a TrMR of $10^5$ was reported~\cite{nature13763}. Unfortunately, the Hall mobility could not be resolved, which leaves the placement of this material Fig.\,\ref{fig4} open (see bar).
Thus, all TrMR values observed to date are consistent with a two-carrier model.

\section{Applicability}

Let us not conclude without mentioning limitations of our analysis  scheme. 
The two-carrier model assumes that the charge carrier concentrations and mobilities, and thus $N$ and $M$ are independent of $B$. 
Systems with a strongly $B$-dependent electronic structure or scattering processes can therefore not be expected to be described.
Whether or not a certain materials class obeys the TrMR limit discussed above is therefore an indication of the validity of these conditions. 
The fact that a large number of Dirac and Weyl fermion systems all conform with the TrMR limit (Fig.\,\ref{fig4}) underpins the validity of this analysis for this materials class. 
By contrast, strongly field-dependent parameters are typically found in strongly correlated electron systems~\cite{nphys3555,PhysRevLett.111.056601}, which may thus break the limit.

Finally we point out that our two-carrier model does not specify the origin of the carriers. 
Thus, the analysis is valid not only for intrinsic two-band transport, but can also describe multi-layer films or extrinsic carriers arising from spatial inhomogeneity.

\section{Summary}
In summary, we have proposed a new robust analysis scheme for magnetotransport due to two independent conduction channels. The scheme is particularly powerful for studies of Dirac of Weyl fermion systems. We showed that the hallmarks of the particularly interesting case of three-dimensional topological insulators (where the highly mobile surface carriers are the majority carriers) are the absence of sign inversion in the Hall coefficient, and huge linear-in-field transverse magnetoresistance values, yet within the universal limit we have established.

\section{Acknowledgement}
We acknowledge fruitful discussions with Kenta Kuroda, Akio Kimura, Yuuichiro Ando, and Masashi Shiraishi, and financial support from the Austrian Science Fund (project FWF I2535-N27). 

\section{Appendix} 
As described in the main part, in the newly proposed analysis scheme of the two-carrier model, the charge carrier types ($q_1$, $q_2$) do not need to be anticipated. 
The four free parameters are transformed into the two new free parameters ($N, M$) and the two fixed parameters ($R_{\rm{H}}, \mu_{\rm{H}}$). The latter two can be read off from the experiments without ambiguity. 
The former two are determined by fitting, which results in much reduced errors compared to the case where four parameters are fit. 
Finally, $q_1$ and $q_2$ are determined as functions of $N$, $M$, and $\mu_{\rm{H}}$. 
Thus, all parameters of the two-carrier model are determined from $R_{\rm{H}}$, $\mu_{\rm{H}}$, $N$, and $M$ with high precision. 
Here, further details of the scheme are presented.

\subsection{Formulation}
\label{formulation}
The common expressions of resistance $R_{xx}(B)$ and Hall resistance $R_{xy}(B)$ in the two-carrier model~\cite{Ashcroft_Mermin} are
\begin{widetext}
\begin{equation}
R_{xx}(B)= \frac{n_1q_1\mu_1 + n_2q_2\mu_2 + (n_1q_1\mu_2+n_2q_2\mu_1)\mu_1\mu_2B^2} {(n_1q_1\mu_1+n_2q_2\mu_2)^2 + (n_1q_1+n_2q_2)^2\mu_1^2\mu_2^2B^2} \quad ,
\label{eq1s}
\end{equation}
and
\begin{equation}
R_{xy}(B)= \frac{n_1q_1\mu_1^2 + n_2q_2\mu_2^2 + (n_1q_1+n_2q_2)\mu_1^2\mu_2^2B^2} {(n_1q_1\mu_1+n_2q_2\mu_2)^2 + (n_1q_1+n_2q_2)^2\mu_1^2\mu_2^2B^2}B \quad .
\label{eq2s}
\end{equation}
In the linear-response regime, $R_{\rm{H}}$ and $\mu_{\rm{H}}$ thus are
\begin{equation}
R_{\rm{H}} = \frac{n_1q_1\mu_1^2 + n_2q_2\mu_2^2} {(n_1q_1\mu_1+n_2q_2\mu_2)^2} \quad ,
\label{eq2s1}
\end{equation}
and 
\begin{equation}
\mu_{\rm{H}} = \frac{n_1q_1\mu_1^2 + n_2q_2\mu_2^2} {n_1q_1\mu_1+n_2q_2\mu_2} \quad .
\label{eq2s2}
\end{equation}
They describe transport in the presence of two independent conduction channels 1 and 2.
For the new analysis scheme these are transformed into
\begin{equation}
R_{xx}(B)= \frac{2}{n_+\mu_+q_1} \Bigg[ \frac{(1+N)(1+M)+(q_2/q_1)(1-N)(1-M)+[(1+N)(1-M)+(q_2/q_1)(1-N)(1+M)](1-M^2)(\mu_+B)^2}{[(1+N)(1+M)+(q_2/q_1)(1-N)(1-M)]^2+[(1+N)+(q_2/q_1)(1-N)]^2(1-M^2)^2(\mu_+B)^2} \Bigg] \quad ,
\label{eq3s}
\end{equation}
and
\begin{equation}
R_{xy}(B)= \frac{2}{n_+q_1} \Bigg[ \frac{(1+N)(1+M)^2+(q_2/q_1)(1-N)(1-M)^2+[(1+N)+(q_2/q_1)(1-N)](1-M^2)^2(\mu_+B)^2}{[(1+N)(1+M)+(q_2/q_1)(1-N)(1-M)]^2+[(1+N)+(q_2/q_1)(1-N)]^2(1-M^2)^2(\mu_+B)^2} \Bigg]B \quad ,
\label{eq4s}
\end{equation}
with $n_+$ and $\mu_+$ being defined as 
\begin{equation}
n_+ \equiv n_1+n_2 = \frac{2}{q_1R_{\rm{H}}} \Bigg[ \frac{(1+N)(1+M)^2+(q_2/q_1)(1-N)(1-M)^2}{[(1+N)(1+M)+(q_2/q_1)(1-N)(1-M)]^2} \Bigg] \quad ,
\label{eq5s}
\end{equation}
and
\begin{equation}
\mu_+ \equiv \frac{\mu_1+\mu_2}{2} = \mu_{\rm{H}} \frac{(1+N)(1+M)+(q_2/q_1)(1-N)(1-M)}{(1+N)(1+M)^2+(q_2/q_1)(1-N)(1-M)^2} \quad .
\label{eq6s}
\end{equation}
\end{widetext}
Therefore, $R_{xx}(B)$ and $R_{xy}(B)$ are characterized by the free parameters $N$ and $M$. 
Finally, all transport properties are defined using $R_{xx}(B)$ and $R_{xy}(B)$. For instance, the transverse magnetoresistance is given by ${\rm{TrMR}}\equiv [R_{xx}(B)-R_{xx}(0)]/R_{xx}(0)$, the Hall angle by ${\rm{tan}} \theta \equiv R_{xy}(B)/R_{xx}(B)$, and the conductance by $S_{xx}(B) \equiv R_{xx}(B)/[(R_{xx}(B)^2+R_{xy}(B)^2]$.
We further define that $n_1$, $q_1$, and $\mu_1$ refer to the majority carriers ($n_1>n_2$) and therefore $0 < N < 1$. 
We have $|M|<1$ for $q_2/q_1=1$ and $|M|>1$ for $q_2/q_1=-1$. 
The ranges of $M$ for each condition are summarized in Table~\ref{tab1}.

\subsection{Determination of carrier types}
The charge carrier types are obtained from $\mu_+$ as given in Eqn.\,\ref{eq6s} and 
\begin{equation}
\mu_- \equiv \frac{\mu_1-\mu_2}{2}  = \mu_+M \quad ,
\label{eq7s}
\end{equation}
which in turn are calculated from $N$, $M$, and $\mu_{\rm{H}}$.
For $q_2/q_1=1$, $q_1$ and $q_2$ are specified by the sign of $\mu_+$. 
For $q_2/q_1=-1$, on the other hand, $q_1$ and $q_2$ are specified by the sign of $\mu_-$. 
This is because the sign is determined from the definition that positive $\mu_-$ values indicate $q_1=e$ and negative $\mu_-$ values indicate $q_1=-e$. 
The ranges of $\mu_\pm$ for each condition are summarized in Table~\ref{tab1}.

\begin{table}[h]
\caption{Ranges of $M$ $\equiv (\mu_1-\mu_2)/(\mu_1+\mu_2)$ and $\mu_\pm$ $\equiv (\mu_1 \pm \mu_2)/2$ for each condition. Here $q_1$, $q_2$ ($\equiv \pm e$) are the charges, $e$ is the elementary charge, and $\mu_1$, $\mu_2$ are the mobilities. $n_1$, $q_1$, and $\mu_1$ refer to the majority carriers ($n_1>n_2$).
Carrier type is determined from the sign of $\mu_\pm$.}
\label{tab1}
\centering
\begin{ruledtabular}
	\begin{tabularx}{1\textwidth}{ccccc} 
		&\multicolumn{1}{c}{Carrier type} & \multicolumn{1}{c}{Mobility} & \multicolumn{1}{c}{$M$} & \multicolumn{1}{c}{$\mu_\pm$} \\ \hline
	 \multirow{4}{*}{$q_2/q_1=1$}  & \multirow{2}{*}{$q_1=-e$} & $|\mu_1| > |\mu_2|$ & $0<M<1$ & \multirow{2}{*}{$\mu_+<0$} \\ 
		&		& $|\mu_1| < |\mu_2|$ & $-1<M<0$  \\ \cline{2-5}
		& \multirow{2}{*}{$q_1=+e$} & $|\mu_1| > |\mu_2|$ & $0<M<1$ & \multirow{2}{*}{$\mu_+>0$} \\ 
		&		& $|\mu_1| < |\mu_2|$ & $-1<M<0$  \\ \cline{1-5}
	\multirow{4}{*}{$q_2/q_1=-1$} & \multirow{2}{*}{$q_1=-e$} & $|\mu_1| > |\mu_2|$ & $1<M<\infty$ & \multirow{2}{*}{$\mu_-<0$} \\ 
		&		& $|\mu_1| < |\mu_2|$ & $-\infty<M<-1$ \\ \cline{2-5}
		& \multirow{2}{*}{$q_1=+e$} & $|\mu_1| > |\mu_2|$ & $1<M<\infty$ & \multirow{2}{*}{$\mu_->0$} \\ 
		&		& $|\mu_1| < |\mu_2|$ & $\infty<M<-1$  \\ 
	\end{tabularx}
\end{ruledtabular}
\end{table}

\begin{figure*}[h]
\begin{center}
\includegraphics[width=\textwidth,clip]{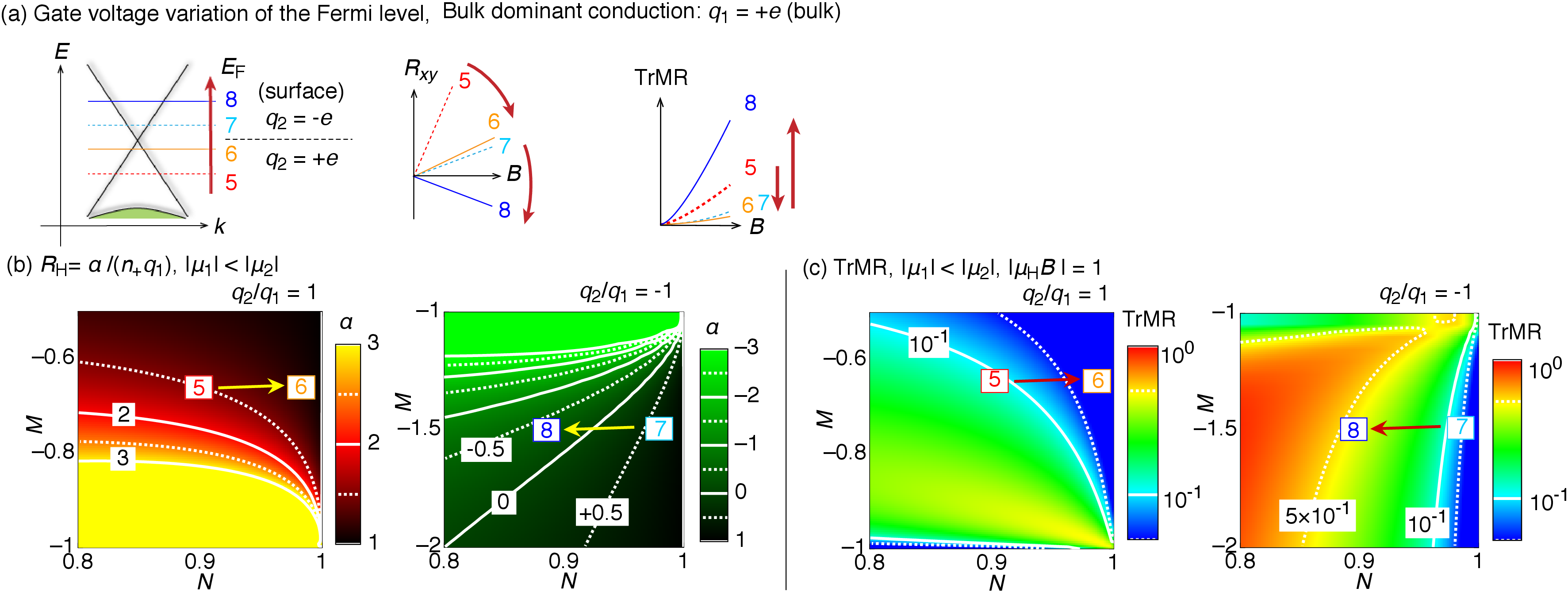}
\caption{(Color online) (a) Sketch of the electronic band dispersion of a 3D-TI around the Dirac point, with an increase of $E_{\rm{F}}$ by $V_{\rm{G}}$ (5: lowest  $E_{\rm{F}}$, 8: highest $E_{\rm{F}}$; left) and the corresponding consecutive $R_{xy}(B)$ (center) and ${\rm{TrMR}}(B)$ (right) traces. 
(b) Contour plots of $\alpha(N,\,M)$ ($q_2/q_1=1$,\,left) around $|\mu_2/\mu_1|=5\,(M=-0.667)$ and $\alpha(N,\,M)$ ($q_2/q_1=-1$,\,right) around $|\mu_2/\mu_1|=5\,(M=-1.5)$.
No sign inversion occurs at the Dirac point ($6\rightarrow 7$). 
(c) Contour plots of ${\rm{TrMR}}(N,\,M,\,|\mu_{\rm{H}}B|=1,\,q_2/q_1=1)$ (left) and ${\rm{TrMR}}(N,\,M,\,|\mu_{\rm{H}}B|=1,\,q_2/q_1=-1)$ (right). 
The minimum TrMR is observed at the Dirac point ($6\rightarrow 7$).}
\label{sfig1s}
\end{center}
\end{figure*}

\subsection{Gate voltage tuning of the Fermi level around the Dirac point}
For $q_2/q_1=1$ and $|\mu_1|<|\mu_2|$, $R_{\rm{H}}$ is transformed from Eqn.\,\ref{eq2s1} into 
\begin{equation}
R_{\rm{H}}= \frac{1}{n_+q_1} \alpha \quad, \quad \quad \alpha= \frac{1+2NM+M^2}{(1+NM)^2} \quad ,
\label{eq7s}
\end{equation}
where $\alpha$ is the Hall factor, and TrMR is transformed from its original form
\begin{equation}
{\rm{TrMR}} = \frac{n_1q_1\mu_1n_2q_2\mu_2(\mu_1-\mu_2)^2B^2}{(n_1q_1\mu_1+n_2q_2\mu_2)^2+(n_1q_1+n_2q_2)^2\mu_1^2\mu_2^2B^2} \quad ,
\label{eq7s0}
\end{equation}
into
\begin{equation}
{\rm{TrMR}} = \frac{(1-N^2)M^2(1-M^2)(\mu_{\rm{H}}B)^2}{(1+2NM+M^2)^2+(1-M^2)^2(\mu_{\rm{H}}B)^2} \quad .
\label{eq8s}
\end{equation}
Under these conditions, $M$ is defined in the range $-1<M<0$. 
Figure~\ref{sfig1s}\,(a) shows the gate voltage $V_{\rm{G}}$ variation around the Dirac point, presented also in the main text (Fig.\,2\,(g)). 
Figures~\ref{sfig1s}\,(b) and (c) are the $\alpha(N,\,M)$ and the TrMR($N,\,M,\,|\mu_{\rm{H}}B|=1$) contour plots, respectively. 
The $R_{xy}(B)$ traces (Fig.\,\ref{sfig1s}\,(a,\,center)) and $\alpha(N,\,M)$ contour plots for $q_2/q_1=1$ (Fermi level $E_{\rm{F}}$ below Dirac point, traces 5 and 6, Fig.\,\ref{sfig1s}\,(b,\,left)) and $q_2/q_1=-1$ ($E_{\rm{F}}$ above Dirac point, traces 7 and 8, Fig.\,\ref{sfig1s}\,(b,\,right)) reveal that charge neutrality occurs for $E_{\rm{F}}$ above the Dirac point. 
The corresponding TrMR traces (Fig.\,\ref{sfig1s}\,(a,\,right)) and contour plots (Fig.\,\ref{sfig1s}\,(c)) reveal that minimal TrMR occurs at the Dirac point ($6\rightarrow 7$). 

\subsection{Application to real systems}
Examples of data analysis are presented in the following. First we explain why only the new analysis scheme can finally, after more than 47 years, overcome the `pitfall of the model' mentioned in \cite{Colin}. Then, using examples from the field of topological materials, we demonstrate how severely the new two-carrier analysis can affect conclusions drawn from the data. Either TrMR or Hall resistivity $\rho_{xy}(B)$ data were examined, depending on which of them were available~\cite{PhysRevB.91.235117,PhysRevB.90.201307,PhysRevB.95.045123,nmat4143,PhysRevX.5.031023}. 

\subsubsection{Topological insulators}
\begin{figure*}[h]
\begin{center}
\includegraphics[width=0.8\textwidth,clip]{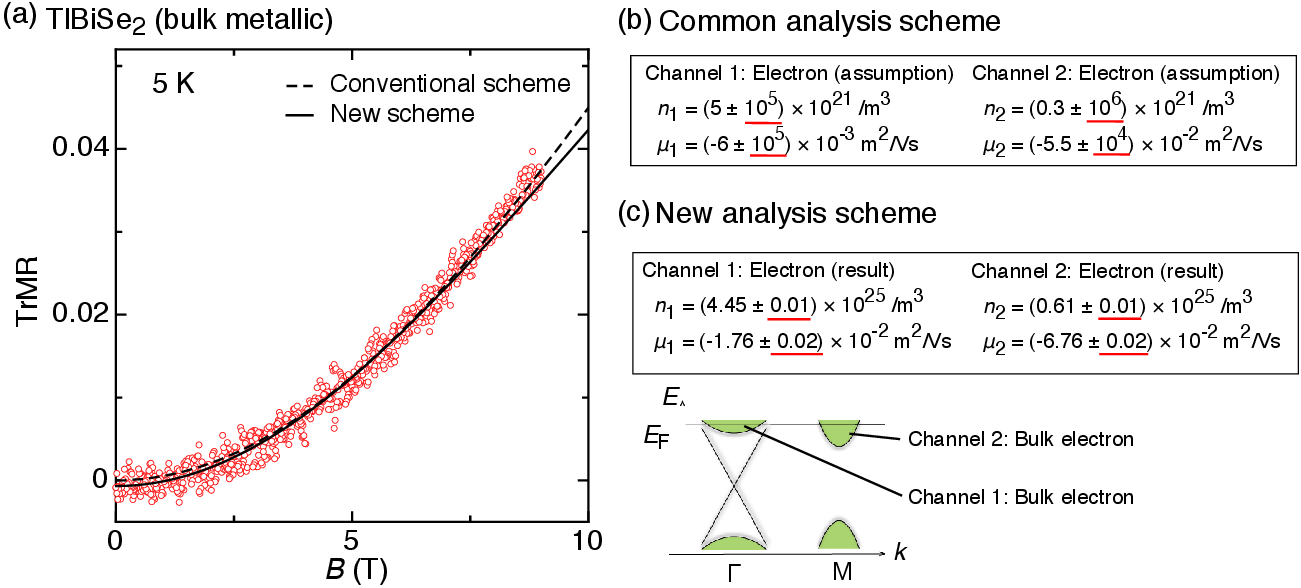}
\caption{(Color online) (a)\,TrMR of a bulk metallic sample of TlBiSe$_2$ presented in Fig.\,3\,(a) of \cite{PhysRevB.91.235117}. The fitting curves resulting from both a common and the new scheme are presented as dashed and full line, respectively. (b)\,Transport parameters obtained from a common analysis scheme. (c)\,Parameters obtained from the new analysis scheme. The schematic shows the electronic band dispersion and $E_{\rm{F}}$ determined from ARPES~\cite{PhysRevB.91.235117}.}
\label{sfig2s1}
\end{center}
\end{figure*}

\begin{figure*}[h]
\begin{center}
\includegraphics[width=0.8\textwidth,clip]{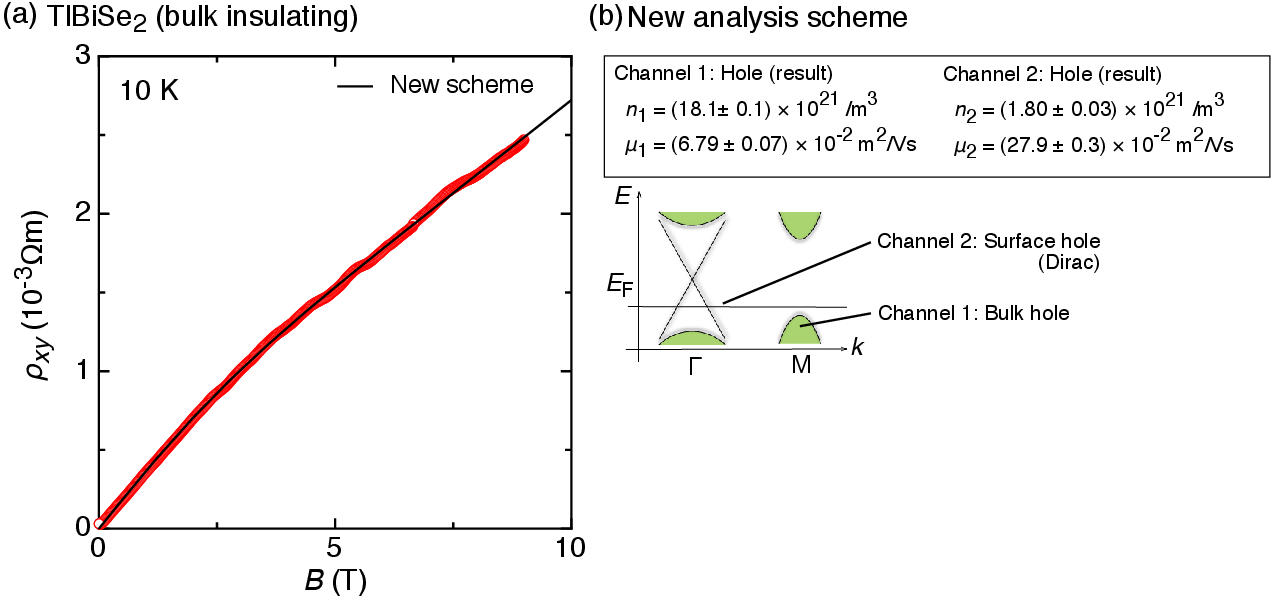}
\caption{(Color online) (a)\,$\rho_{xy}(B)$ of a bulk insulating sample of TlBiSe$_2$ presented in Fig.\,1\,(e) of \cite{PhysRevB.90.201307}. The fitting curve resulting from the new scheme is presented as full line. (b)\,Transport parameters obtained from the new analysis scheme. The errors are all below 2\%, making the results highly meaningful. The schematic shows the electronic band dispersion~\cite{PhysRevB.91.235117,PhysRevB.90.201307} and the position of $E_{\rm{F}}$ extracted from the analysis. 
}
\label{sfig2s2}
\end{center}
\end{figure*}

\begin{figure*}[h]
\begin{center}
\includegraphics[width=0.8\textwidth,clip]{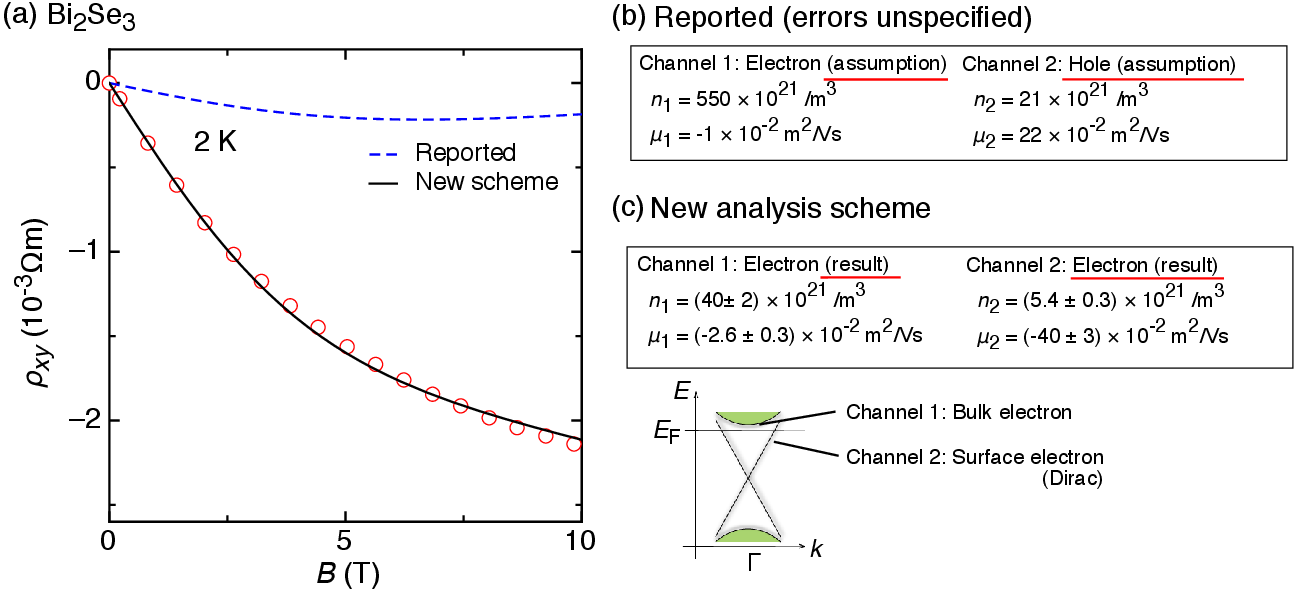}
\caption{(Color online)  (a)\,$\rho_{xy}(B)$ of Bi$_2$Se$_3$ digitized from Fig.\,3\,(d, \,sample A) of \cite{PhysRevB.95.045123}. The curves resulting from the reported parameters and from the fitting using the new scheme are presented as dashed and full line, respectively. (b)\,Transport parameters reported in \cite{PhysRevB.95.045123}. The errors are not available from the literature. (c)\,Parameters obtained from the new analysis scheme. The errors are all below 12\%, making the results meaningful. The schematic shows the electronic band dispersion~\cite{PhysRevB.95.045123} and the $E_{\rm{F}}$ extracted from the analysis.
}
\label{sfig2s4}
\end{center}
\end{figure*}

The first example is the topological insulator TlBiSe$_2$, with $E_{\rm{F}}$ in the bulk conduction band~\cite{PhysRevB.91.235117}. Figure~\ref{sfig2s1}\,(a) shows TrMR data and fitting curves determined by common and new analysis schemes. In the former case errors are as large as 10$^7$\%, which indicates that the results (Fig.\,\ref{sfig2s1}\,(b)) are meaningless. By contrast, in the latter case the errors are at most 1.7\%, making the results (Fig.\,\ref{sfig2s1}\,(c)) highly meaningful. Note that the latter agrees well with the electronic band dispersion determined by angle resolved photoemission spectroscopy (ARPES)~\cite{PhysRevB.91.235117}. 

The second example is TlBiSe$_2$ with $E_{\rm{F}}$ in the bulk band gap~\cite{PhysRevB.90.201307}, for which previously no two-carrier analysis has been performed. The new analysis scheme does not require assumption of carrier types and, surprisingly, it revealed conduction with two hole channels\,(Fig.\,\ref{sfig2s2}). Note that the obtained values for channel 2 agree well with those detected in Shubnikov de-Hass (SdH) measurements~\cite{PhysRevB.90.201307}.

The third example is the nonmetallic stoichiometric topological insulator Bi$_2$Se$_3$\,(Fig.\,\ref{sfig2s4})~\cite{PhysRevB.95.045123}. The published two-carrier analysis assumes two electron channels for one sample (sample V), and one electron and one hole channel for another sample (sample A), with electrons being the bulk carriers in both cases~\cite{PhysRevB.95.045123}. Based on these analyses, the authors make the interesting claim that in sample A surface Dirac holes are present, without any gating. However, as shown in Fig.\,\ref{sfig2s4}\,(a), the reported parameters of sample A (Fig.\,3(e,\,f) of~\cite{PhysRevB.95.045123}) do not reproduce the corresponding data (Fig.\,3\,(d) of~\cite{PhysRevB.95.045123}). We therefore reanalyzed the $\rho_{xy}(B)$ data with our new scheme. Surprisingly, the analysis reveals conduction with two electron channels instead of one electron and one hole channel, meaning that also the Dirac surface carriers are electron like. This is fully consistent with the SdH measurements also presented in~\cite{PhysRevB.95.045123}, which cannot determine the carrier types. Thus, our new scheme shows that the key message of that work needs to be revised.

\subsubsection{Dirac and Weyl semimetals}
\begin{figure*}[h]
\begin{center}
\includegraphics[width=0.8\textwidth,clip]{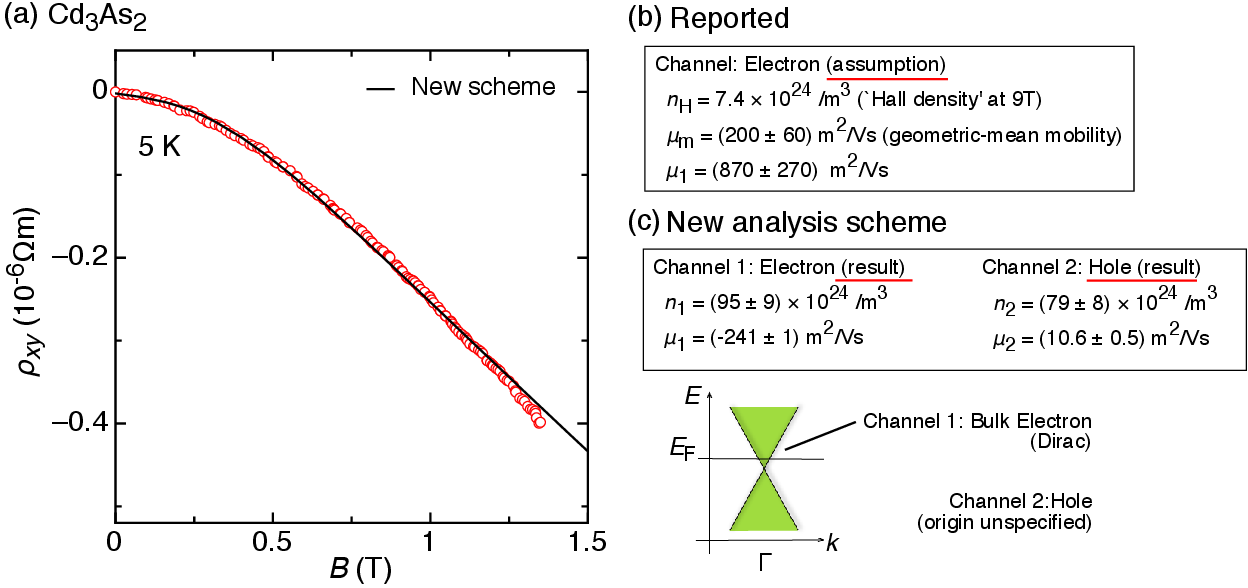}
\caption{(Color online) (a)\,$\rho_{xy}(B)$ of Cd$_3$As$_2$ digitized from Fig.\,2\,(c) of \cite{nmat4143}. The fitting curve resulting from the new scheme is presented as full line. (b)\,Transport parameters reported in \cite{nmat4143}. (c)\,Parameters obtained from the new analysis scheme. The errors are all below 10\%, making the results meaningful. The schematic shows the electronic band dispersion~\cite{nmat4143} and $E_{\rm{F}}$ extracted from the analysis. The origin of the hole channel remains to be clarified.}
\label{sfig2s5}
\end{center}
\end{figure*}

\begin{figure*}[h]
\begin{center}
\includegraphics[width=0.8\textwidth,clip]{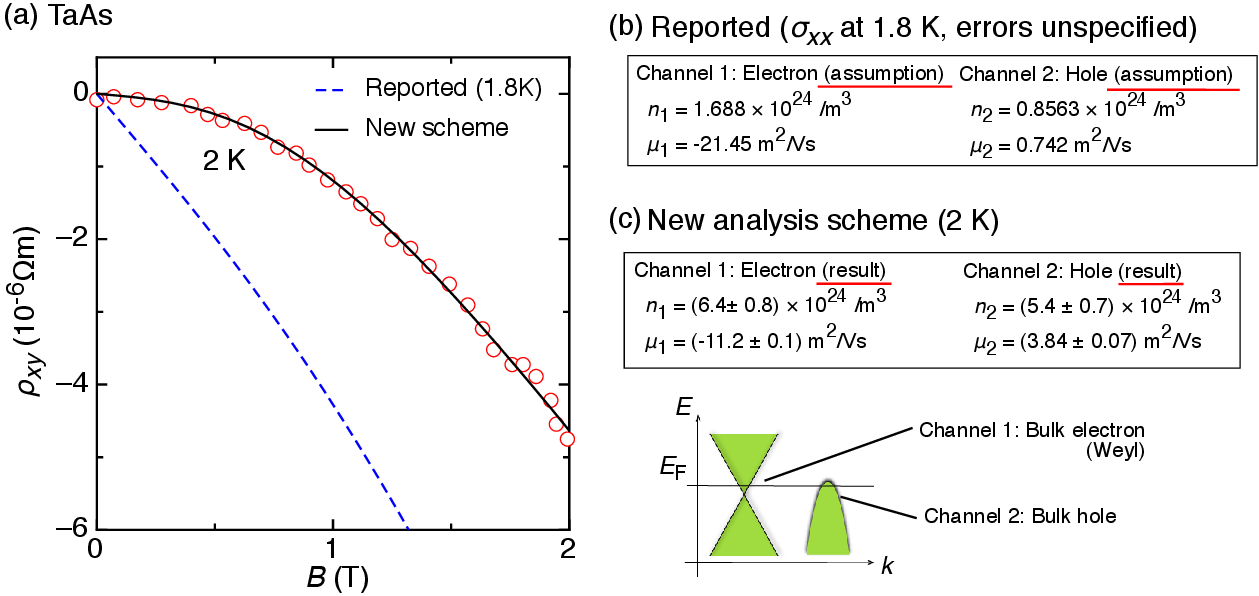}
\caption{(Color online) (a)\,$\rho_{xy}(B)$ of TaAs digitized from Fig.\,3\,(a) inset of \cite{PhysRevX.5.031023}. The curves resulting from the reported parameters and from the fitting using the new scheme are presented as dashed and full line, respectively. (b)\,Transport parameters reported in \cite{PhysRevX.5.031023}. The errors are not available from the literature. (c)\,Parameters obtained from the new analysis scheme. The errors are all below 13\%, making the results meaningful. The schematic shows the electronic band dispersion~\cite{PhysRevX.5.031023} and $E_{\rm{F}}$ extracted from the analysis.
}
\label{sfig2s6}
\end{center}
\end{figure*}

As demonstrated above, analyses of TrMR and non-linear $\rho_{xy}$ depend largely on the pre-assumed carrier types. Note that, within the single-carrier model the TrMR and the non-linear $\rho_{xy}$ are absent, and for this reason the two-carrier model is recognized to be the standard model for their analysis~\cite{Ashcroft_Mermin,Colin}. In the following we further provide analyses of the Dirac semimetal Cd$_3$As$_2$\,(Fig.\,\ref{sfig2s5})~\cite{nmat4143} and of the Weyl semimetal TaAs\,(Fig.\,\ref{sfig2s6})~\cite{PhysRevX.5.031023}.

In \cite{nmat4143} ultrahigh mobility and giant TrMR in Ce$_3$As$_2$ are reported, and the authors attribute them to conduction by a single topologically-protected electron channel with extremely long transport lifetime. We reanalyzed the $\rho_{xy}(B)$ data with our new scheme and find that there are two channels instead: 55\% electrons with significantly degraded lifetime, and 45\% holes. Our new analysis also gives a natural interpretation of the giant observed TrMR signal. Whereas it was previously attributed to anisotropy in the single Fermi pocket, we here put forward a completely different mechanism, namely archetypal behavior originating from compensated two-channel conduction. 

In \cite{PhysRevX.5.031023} negative longitudinal magnetoresistance in TaAs is reported and, based on the authors' two-carrier analysis, attributed to the chirality of predominant Weyl electrons. Figure~\ref{sfig2s6}\,(a) shows the $\rho_{xy}(B)$ data and the curve derived from the reported parameters (broken line). It clearly fails to reproduce the data, just in the same way as shown in Fig.\,\ref{sfig2s4}\,(a). Our new analysis (solid line) reveals a largely enhanced transport contribution from an additional hole channel, that has been considered as topologically trivial\,\cite{PhysRevX.5.031023}. Thus, the dominance of Weyl electrons is much less drastic than claimed, calling into question their relation with the observed negative longitudinal magnetoresistance.

\bibliographystyle{apsrev4-1}
\bibliography{Preload,myrefs}

\end{document}